\documentclass[aps,showpacs]{revtex4}

\usepackage{natbib}
\usepackage{epsfig}
\usepackage{graphicx}
\usepackage{multirow}
\usepackage{amssymb}
\usepackage{amsfonts}

\begin{document}

\title{Dose distribution in water for monoenergetic photon point
sources in the energy range of interest in brachytherapy: Monte Carlo
simulations with PENELOPE and GEANT4}

\author{Julio F. Almansa$ ^{\,1}$, Rafael Guerrero$ ^{\,2}$, 
Feras M.O. Al-Dweri$ ^{\,3}$, M. Anguiano$ ^{\,3}$ and
A.M. Lallena$^{\,3}$}

\affiliation{$^{1)}$ Servicio de Radiof\'{\i}sica y Protecci\'on 
Radiol\'ogica, Hospital Universitario ``Puerta del Mar'', E-11009
C\'adiz, Spain.\\
$^{2)}$ Servicio de Radiof\'{\i}sica, 
Hospital Universitario ``San Cecilio'',
Avda. Dr. Ol\'oriz, 16, E-18012 Granada, Spain.\\
$^{3)}$ Departamento de F\'{\i}sica At\'omica, Molecular y Nuclear,
Universidad de Granada, E-18071 Granada, Spain.}

\begin{abstract}
Monte Carlo calculations using the codes PENELOPE and GEANT4 have been
performed to characterize the dosimetric properties of monoenergetic
photon point sources in water. The dose rate in water has been
calculated for energies of interest in brachytherapy, ranging between
10~keV and 2~MeV. A comparison of the results obtained using the two
codes with the available data calculated with other Monte Carlo codes
is carried out. A $\chi^2$-like statistical test is proposed for these
comparisons. PENELOPE and GEANT4 show a reasonable agreement for all
energies analyzed and distances to the source larger than
1~cm. Significant differences are found at distances from the source
up to 1~cm. A similar situation occurs between PENELOPE and EGS4.
\end{abstract}


\maketitle

\section{Introduction}

Dosimetry of monoenergetic point sources is interesting from different
points of view. Dose distributions around more complex sources
(e.g. polyenergetic sources or sources with finite dimensions) can be
calculated using the monochromatic source data, providing the
corresponding modifications to account for the source geometry and the
capsule materials are made.

Dose distributions generated by monoenergetic point sources constitute
a crucial test to compare Monte Carlo (MC) radiation transport codes,
because the differences observed in the simulation results can be
linked to the basic physics taken into account by these codes.

In case of photons, the most complete available databases of dose
distributions in water for monoenergetic point sources are those of
\citet{Ber68} and \citet{Lux99}. Berger solved the one-dimensional
Boltzmann transport equation and showed results for energies ranging
between 15~keV and 3~MeV. He provided polynomial expansions of the
energy-absorption buildup factors needed to calculate the
corresponding dose rates.

More recently, \citet{Lux99} have used the EGS4 MC code to
determine these dose distributions for 10~keV to 2~MeV photons. These
authors have found a good agreement with Berger's results (within 1\%)
between 40 and 400~keV. Some discrepancies have been quoted for
energies below 40~keV, where differences reach 4-5\%. Above 500~keV,
Luxton and Jozsef have found a buildup in the radial dose distribution
not shown in the results of Berger. Between 1 and 2~MeV, discrepancies
are $\sim$4\%. Finally, EGS4 results suggest an additional structure
at $\sim$3~cm from the 2~MeV source, which the authors have guessed to
be an artifact of the MC simulation.

The MC code PENELOPE (v. 2001) \citep{Sal01} has been benchmarked for
low-energy photons (10-150~keV) by \citet{Ye04}. These authors have
compared the doses calculated with PENELOPE and MCNP4C to the results
of \citet{Lux99} obtained with EGS4. They paid special attention to
the differences due to the various libraries used: EPDL97
\citep{Cul97} for PENELOPE, DLC-200 \citep{Fra00} and the updated
DLC-146 for MCNP4C and PHOTX \citep{Sak93} for EGS4. Ye et al. found
that all the results agreed rather well except those obtained with
MCNP4C and the DLC200 library below 100~keV for which the dose rates
are up to 9\% lower than for the other cases.

In their MC simulations, \citet{Ye04} used a cylindrical
liquid water phantom with 30~cm diameter. For energies
above 100~keV, Ye et al. observed discrepancies between their
calculations and those of \citet{Lux99} which they ascribed
to the fact that these authors used an infinite water phantom.

In this work, new simulations of monoenergetic photon point sources in
the range 10~keV to 2~MeV embedded in water are performed using the MC
codes PENELOPE \citep{Sal01,Sal03} and GEANT4
\citep{Ago03,GEANT4}. Our purpose is to benchmark the PENELOPE and
GEANT4 codes for this energy range by intercomparing our results with
those found by \citet{Lux99}, with EGS4, and by \citet{Ye04}, also
with PENELOPE but with a different scoring geometry.

\section{Material and Methods}

\subsection{Monte Carlo codes}

In this work PENELOPE (v. 2001) \citep{Sal01} and (v. 2003)
\citep{Sal03} and GEANT4 \citep{Ago03,GEANT4}
(v. 4.5.2) MC codes have been used to perform the simulations.

PENELOPE is a general purpose MC code which performs simulations of
coupled electron-photon transport. It can be applied for energies
ranging from a few hundred eV up to 1~GeV and for arbitrary
materials. Besides, PENELOPE permits a good description of the
particle transport at the interfaces and presents a more accurate
description of the electron transport at low energies in comparison to
other general purpose MC codes. These characteristics make PENELOPE to
be an useful tool for medical physics applications as previous works
have pointed out \citep[see e.g.][]{San98,Ase02,Ald04}. Details about
the physical processes considered can be found in \citet{Sal01,Sal03}.

GEANT4 is a toolkit for simulating the passage of particles
through matter for applications in high energy physics, nuclear
experiments, medical physics studies, etc. Due to its purpose, the
available energy range and the possible particles to be simulated are
much larger than in PENELOPE. GEANT4 includes low energy packages
which permit to extend the validity range of particle interactions for
electrons, positrons and photons down to 250 eV, and can be used up to
approximately 100 GeV. A general description of the GEANT4 toolkit can
be found in \citet{Ago03}. A detailed documentation
concerning the physics involved can be found in \citet{GEANT4}. In
our calculations the low-energy package G4EMLOW.2.3 has been used.

\subsection{Simulations}
\label{sec:phantoms}

In the simulations performed we have used an infinite water phantom
which we have labeled $\mathcal{P}_{\rm inf}$. In this case the
scoring voxels have been chosen to be concentric spherical shells with
thickness 0.5 mm and centered in the point source. The size of these
scoring voxels avoids the possible volume averaging artifacts that
could appear using larger voxels.

Besides, we have performed additional simulations with a spherical
finite phantom with a radius of 15~cm, filled with water and with the
point source situated at its center. We have labeled $\mathcal{P}_{\rm
fin}$ this phantom and we have used the same scoring voxels as in
$\mathcal{P}_{\rm inf}$. 

As the sources studied are point sources, the different quantities
simulated (e.g., the dose rate, $\dot{D}(r)$), depend on the distance
between the target point and the source, $r$.

In our simulations, the energy deposited in each voxel $j$, $E^j$, was
scored to obtain the histograms corresponding to $\dot{D}(r)$. We
labeled $\dot{D}^j$ the value corresponding to the $j$-th voxel. In
order to avoid the effect of the geometric factor ($r^{-2}$ in this
case) which dominates $\dot{D}(r)$, we have calculated also the
quantity
\begin{equation}
R(r)\, = \, r^2 \, \dot{D}(r) \, .
\label{eq:r2D}
\end{equation}
We have simulated this quantity by scoring directly $(r^2 E)$, thus
obtaining $R^j$ for each scoring shell $j$. However, in common
practice (see e.g. Luxton and Jozsef 1999), this quantity is
calculated as
\begin{equation}
R^j_{\rm M} \, = \, (r^j_{\rm M})^2 \, \dot{D}^j \, ,
\label{eq:rM2D}
\end{equation}
for each bin $j$ of the histogram corresponding to $\dot{D}(r)$. In
this equation 
\begin{equation}
r^j_{\rm M} \, = \, \frac{r^j_>+r^j_<}{2} \, ,
\label{eq:rM}
\end{equation}
with $r^j_<$ and $r^j_>$ the minimum and maximum $r$ values
corresponding to the $j$-th ``detector'' voxel.

The results for $R(r)$ obtained with the phantoms $\mathcal{P}_{\rm
inf}$ and $\mathcal{P}_{\rm fin}$ can be found in {\tt
http://fm137.ugr.es/PhotonPointSources/} for both PENELOPE (v. 2003)
and GEANT4. A total of 40 energies in the range between 10~keV and
2~MeV have been considered. In these tables, also the air kerma
strengths and the dose rate constants calculated with PENELOPE and
GEANT4 are given.

\subsection{Simulation parameters}

PENELOPE considers analog simulation for photons. Electrons and
positrons are simulated by means of a mixed scheme in which collisions
are classified as ``hard'' and ``soft''. Hard collisions are
characterized by a scattering angle or an energy loss larger than
certain cutoff values and are individually simulated. A multiple
scattering theory is used to describe the soft collisions, in which
the polar angular deflection and the energy loss are below the
threshold values. The electron tracking is controlled by means of
four parameters. $C_1$ and $C_2$ refer to elastic collisions. $C_1$
gives the average angular deflection due to a elastic hard collision
and to the soft collisions previous to it. $C_2$ represents the
maximum value permitted for the average fractional energy loss in a
step. On the other hand, $W_{\rm cc}$ and $W_{\rm cr}$ are energy
cutoffs to distinguish hard and soft events. Thus, the inelastic
electron collisions with energy loss $W<W_{\rm cc}$ and the emission
of bremsstrahlung photons with energy $W<W_{\rm cr}$ are considered in
the simulation as soft interactions. In addition the maximum step size
can be controlled using the parameter $s_{\rm max}$.

In our simulations with PENELOPE, the parameters were fixed to the
following values: $W_{\rm cc}=10$~keV, $W_{\rm cr}=1$~keV,
$C_1=C_2=0.05$. Photons were simulated down to 1~keV. Electrons and
positrons were absorbed when they slow down to kinetic energies of
10~keV. We have checked that the use of lower values for these
parameters does not cause changes in the results. Finally, $s_{\rm
max}$ was taken to be $10^{35}$~cm.

In case of GEANT4 simulations, it is necessary to choose a range
threshold for each particle type. This threshold is converted,
internally, to an energy threshold below which secondary particles are
not emitted. In our calculations the range thresholds have been fixed
to values providing energy thresholds of 1 keV and 10 keV for photons
and electrons, respectively. Besides, one can select for electrons the
values of the {\it maximum step size}, the variable $dRoverRange$,
which is the maximum fraction of the stopping range that an electron
can travel in a step, the variable $finalRange$, which establishes the
limit for the gradual reduction of the step size, and, finally, the
variable $f_r$, which controls the step size when the electron is
transported away from a boundary into a new volume. In our simulations
we have used the default values for these parameters (no maximum step
size, $dRoverRange=1$, $finalRange=1$~mm, and $f_r$=0.2). We have
performed a series of calculations using $dRoverRange=0.01$ and
$finalRange=100$~nm. These values were used by \citet{Poo05a} to
investigate the consistency of GEANT4 in geometries including
ionization chambers, which is a more exigent situation than ours
concerning the electron transport. However, we have not found any
difference between the results obtained with this new set of
parameters and the previous one, in the energy range we have
considered.

\subsection{Statistics}

In the simulations we scored the energy deposited by all the particles
of the $i$-th history (that is, including the primary particle and all
the secondaries it generates) in the $j$-th voxel. We label the value
obtained $E^j_i$. This permits to obtain the MC estimate of the
average value of $E^j$ (per initial particle), which is given by
\begin{equation}
 \overline{E}^j \, = \, \frac{1}{N} \sum_{i=1}^{N} E^j_{i} \, .
\end{equation}
Here $N$ is the number of simulated histories. In addition,
$(E^j_{i})^2$ was also scored in order to calculate the statistical
uncertainties which are given by
\begin{equation}
 \sigma_{\overline{E}^j} \, = \, 
 \sqrt{\frac{1}{N} \left[ \frac{1}{N}\sum_{i=1}^{N} (E^j_{i})^2 
\, - \, (\overline{E}^j)^2 \right] } \, .
\end{equation}
We have proceed in a similar way with $(r^2 E)^j$. The uncertainties
given below correspond to 1$\sigma$. Except when explicitly
mentioned, all simulations have been performed following $3 \cdot
10^7$ histories. This permitted us to maintain these statistical
uncertainties under reasonable levels.

\subsection{Comparison of histograms}
\label{sec:histos}

The usual outputs of the MC simulations of radioactive sources include
histograms corresponding to the magnitudes of interest. Then, the
benchmark of the codes used is done by comparing either the histograms
{\it by-eye} in a figure or a restricted set of values corresponding
to different bins in a table.

To gain more insight, it is useful to plot the relative difference
between the two histograms $H_1$ and $H_2$, which is given by
\begin{equation}
\Delta^j_{H_1,H_2} \, = \, \displaystyle  
\frac{H_1^j - H_2^j}{H_1^j} \, ,
\label{eq:reldiff}
\end{equation}
for the different bins $j$.

\begin{figure}[ht]
\begin{center}
\parbox[c]{15cm}
{\includegraphics[scale=0.6]{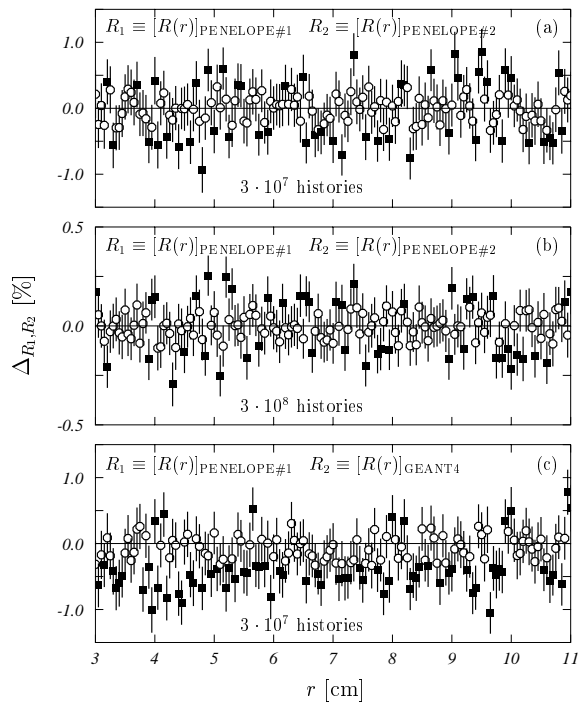}}
\end{center}
\vspace*{-.5cm}
\caption{\small Relative differences in percentage, as given by
  Eq. (\ref{eq:reldiff}), between the values of the quantity $R(r)$
  obtained in different calculations, as a function of the distance to
  the source $r$ and for 80 keV photons. Panels (a) and (b) correspond
  to the two calculations performed with PENELOPE for $3\cdot 10^7$
  and $3\cdot 10^8$ histories, respectively. Panel (c) corresponds to
  the difference between the first calculation with PENELOPE and the
  GEANT4 one for $3\cdot 10^7$ histories. Open circles (black squares)
  represent the results for which the value zero is (not)
  statistically compatible at the 1$\sigma$ level.
\label{fig:histos}}
\end{figure}

To illustrate a typical situation, we show in Fig. \ref{fig:histos},
the relative differences, $\Delta^j_{R_1,R_2}$, in percentage, between
the quantities $R(r)$ obtained in various calculations, as a function
of $r$ and for 80 keV photons. Open circles (black squares) correspond
to the bins for which the zero difference is (not) reached at the
1$\sigma$ level.

In panel (a) the two simulations compared have been performed with
PENELOPE (v. 2003), but using two different seeds for the random
number generator. A total of $3\cdot 10^7$ histories have been
followed in this case. It is obvious that these two simulations must
be identical. However, as we can see, there are some bins for which
both calculations are compatible, but some of them are statistically
different, even if the differences are really small (below $\sim$1\%).
Then, it is evident that, depending on the particular subset of bins
selected, one could conclude that the agreement between the two
calculations is almost perfect or exactly the opposite.

Increasing the number of histories does not solve the problem. In
panel (b) we show the relative differences obtained in a similar
situation to that plotted in panel (a), but following $3\cdot 10^8$
histories. We observe that the general trend is the same in both
panels, the only difference being that now the values of
$\Delta^j_{R_1,R_2}$ are smaller when the number of histories
increases. But, again, it is still possible to select a subset of
individual bins for which the two simulations appear to be
statistically different or very similar. Same situation is found if,
instead of increasing the number of histories, the voxel volume is
enhanced.

Finally, in panel (c) we compare two simulations done with PENELOPE
(v. 2003) and GEANT4 for $3\cdot 10^7$ histories. Despite the fact
that, now, two really different simulations are compared, the values
obtained for $\Delta^j_{R_1,R_2}$ are similar to those plotted in
panel (a) and, as in the previous cases, one should conclude that both
calculations are identical or completely different by adequately
choosing particular bins for the comparison. However, in this
particular case there is an additional fact to be taken into account:
we observe a certain bias pointed out by the presence of a larger
number of black squares with a negative value of the relative
difference.

In order to avoid the possible errors linked to a restricted
comparison between histograms, we propose to use a $\chi^2$-type
statistic defined as
\begin{equation}
\chi^2_{H_1,H_2} \, = \, \displaystyle \sum_{j=1}^M 
\frac{(H_1^j - H_2^j)^2}{(\sigma_1^j)^2 + (\sigma_2^j)^2} \, ,
\label{eq:chi2statistic}
\end{equation}
where $H_1^j$ and $H_2^j$ label the value in the $j$-th bin of the
first and second histograms, respectively, and $\sigma_1^j$ and
$\sigma_2^j$ are the corresponding uncertainties. One can state that
two histograms are similar if $\chi^2_{H_1,H_2}/M \sim 1$, $M$ being
the number of degrees-of freedom (the number of histogram bins in our
case). A value of $\chi^2_{H_1,H_2}/M$ below 1 points out the presence
of correlations between the quantities scored in the histograms. On
the other hand, a value of $\chi^2_{H_1,H_2}/M > 1$ would indicate a
discrepancy between the histograms \citep{Fro79,Pre92}. The uncertainty
of the quantity $\chi^2$ is given by $\sqrt{2\chi^2}$ \citep{Fro79}. In
the three cases considered in Fig. \ref{fig:histos}, the values of
$\chi^2/M$ are 0.96$\pm$0.04, 0.97$\pm$0.04 and 1.39$\pm$0.05,
respectively.

Besides, to estimate the significance of this $\chi^2$ test, we have
used the standard chi-square probability function $Q(\chi^2 | M)$
\citep{Pre92}.  The smaller this probability is, the bigger the
differences between the two histograms are. Following \citet{Pre92} we
can state that if $Q \geq 0.1$ the agreement between both histograms
is believable. If $0.001 \leq Q < 0.1$ the agreement is acceptable.
Finally, if $Q < 0.001$ both histograms have significant differences.
In the two first cases shown in Fig. \ref{fig:histos}, the values of
$Q(\chi^2 | M)$ are 0.807 and 0.755, respectively. In the case of the
comparison between PENELOPE and GEANT4 (panel (c)) this probability is
2.88$\cdot 10^{-15}$, much smaller than 0.001.

\section{Results}

\subsection{Comparison between PENELOPE (v. 2001) and (v. 2003).}

\begin{figure}[b]
\begin{center}
\parbox[c]{15cm}
{\includegraphics[scale=0.6]{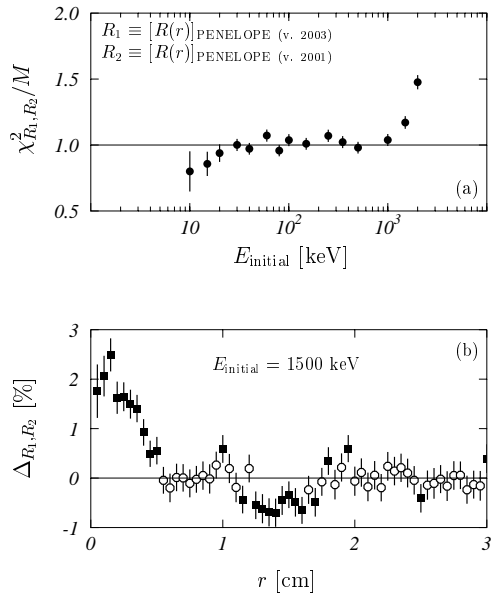}}
\end{center}
\vspace*{-.5cm}
\caption{\small Panel (a): Values of $\chi^2_{R_1,R_2}/M$, as given by
  Eq. (\ref{eq:chi2statistic}), calculated for the quantity $R(r)$, as
  a function of the energy of the initial photons.  Panel (b):
  Relative differences in percentage, as given by
  Eq. (\ref{eq:reldiff}), calculated for the same quantity, as a
  function of the distance to the source $r$ and for photons with
  initial energy of 1500 keV. The calculations compared have been done
  with the two versions of PENELOPE. The meaning of the open circles
  and the black squares in panel (b) is the same as in
  Fig. \ref{fig:histos}.
\label{fig:chi-01-03}}
\end{figure}

Before analyzing the obtained results, it is worth to point out that
we have found small differences between the calculations performed
with the two versions of PENELOPE. These differences appear mainly for
the larger energies considered.  This can be seen in panel (a) of
Fig. \ref{fig:chi-01-03} where the values of the $\chi^2_{R_1,R_2}/M$,
obtained for calculations done with PENELOPE versions 2003 ($R_1$) and 
2001 ($R_2$) and with the phantom $\mathcal{P}_{\rm inf}$,
are shown as a function of the initial photon energy. As we can see,
the values of the quantity $\chi^2_{R_1,R_2}/M$ obtained are close to
1, thus indicating the statistical consistency between the
calculations performed with both versions of the code. The
probabilities $Q(\chi^2 | M)$ obtained are above 0.1 for all energies
between 10~keV and 1~MeV, except for 60 and 250 keV where this
probability is 0.06 in both cases. Above 1~MeV the two histograms show
significant differences, the probabilities found being smaller than
0.001. In panel (b) of Fig. \ref{fig:chi-01-03} we have plotted the
relative differences, as given by Eq. (\ref{eq:reldiff}), between the
values of the quantity $R(r)$ obtained with the two versions of
PENELOPE for an initial energy of 1.5~MeV. Therein the meaning of
black squares and open circles is the same as in
Fig. \ref{fig:histos}. As can be seen the discrepancy is mainly due to
the first few bins of the corresponding histograms. In fact, if one
does not consider the first bins for which the relative differences
are larger than 1.5\%, a probability $Q(\chi^2 | M) > 0.001$ is
obtained.

The reason for these differences between both versions of the MC code
PENELOPE can be ascribed to the modifications introduced in the
version 2003 with respect to the 2001 one \citep{Sal01,Sal03}: 
the X-ray energies are the experimental ones taken from \citet{Bea67}
(instead of being calculated as ionization energy differences as in
the version 2001), the generalized oscillator strength model is fixed in
order to reproduce the stopping powers quoted in \citet{ICRU}
and the inner-shell ionization by electron and positron impact is an
independent process (which was not included in the version 2001). In
what follows we use the version 2003 of PENELOPE only.

A few simulations performed, for some of the cases considered here, with 
the new versions 2005 and 2006 of the code have produced the same results 
as the version 2003.

\subsection{Radial dose rate distributions}

\begin{figure}[b]
\begin{center}
\parbox[c]{15cm}
{\includegraphics[scale=0.6]{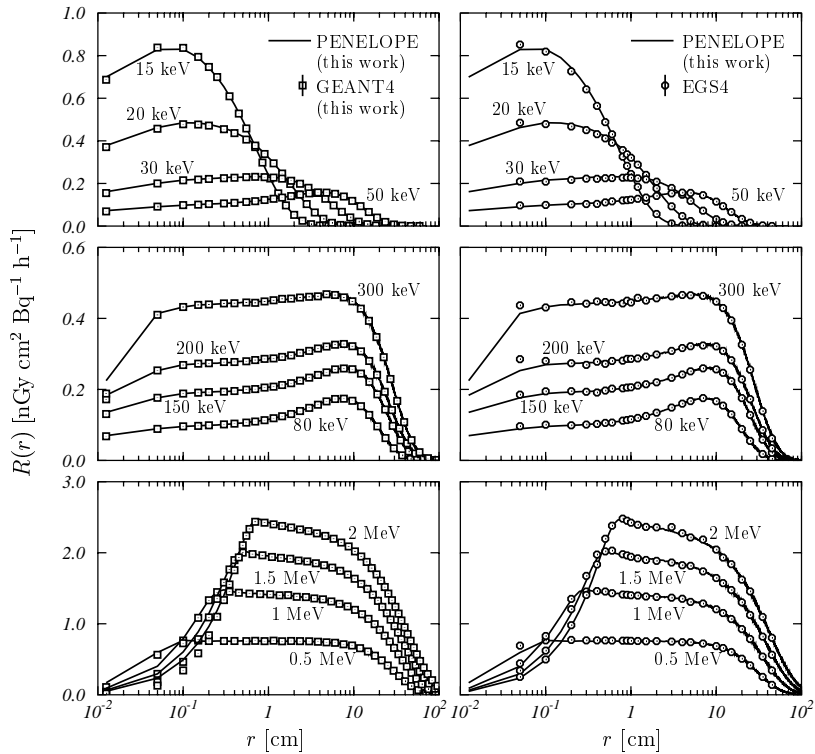}}
\end{center}
\vspace*{-.5cm}
\caption{\small Comparison of the values of $R(r)$, as given by
  Eq. (\ref{eq:rM2D}), obtained with PENELOPE (solid curves)
  with those calculated with GEANT4 (left panels) and with the results
  reported by \citet{Lux99} with EGS4 (right
  panels). Various initial photon energies ranging from 15 keV to 2
  MeV are plotted. Our calculations have been done with the 
  ${\mathcal P}_{\rm inf}$ phantom. The uncertainties for the GEANT4
  calculations are smaller than the  size of the symbols used and those
  of the PENELOPE calculations are of the same order.
\label{fig:comp1}}
\end{figure}

In this section we compare the results of our simulations with those
obtained by \citet{Lux99} and \citet{Ye04}. Fig. \ref{fig:comp1} shows
the values of $R(r)$, as given by Eq. (\ref{eq:rM2D}) for energies
ranging between 15~keV and 2~MeV, as a function of the distance to the
source, $r$, in the ${\mathcal P}_{\rm inf}$ phantom. In all panels,
solid curves correspond to the simulations we have performed with
PENELOPE. These are compared with the results obtained for GEANT4
(open squares) in left panels and with the results of the EGS4
simulation performed by \citet{Lux99} (open circles) in right
panels. An uncertainty of 1\% was assumed for the data of these
authors.

The results of PENELOPE and GEANT4 calculations agree rather well
except at distances to the source smaller than 1~cm. Above 1~cm and
for all energies, the differences between both calculations are below
1\% of the maximum value of the corresponding $R(r)$. 

At distances to the source smaller than 1~cm, the differences are, in
general, larger. Below 0.5~cm and for energies smaller than 80~keV,
the differences are bigger than 1\% of the maximum of $R(r)$, reaching
values around 3\% of this maximum for the first bin (from 0 to
0.25~mm). In the energy range between 80 and 300 keV, only the first
bin gives differences larger than 1\% of the maximum value of
$R(r)$. Above 300~keV, the differences are bigger than 2\% of the
maximum of $R(r)$ for distances to the source smaller than 0.5~cm. For
the first bins, values around 8\% of the maximum of $R(r)$ are found.

The differences between PENELOPE and EGS4 (Luxton and Jozsef 1999)
follow the same trends for all the energies analyzed. At distances to
the source above 1~cm, these differences are below 2\% of the maximum
of the corresponding $R(r)$. For distances smaller than 1~cm, the
differences are at most 6\% of the maximum of $R(r)$. Nevertheless,
the first point available in the work of \citet{Lux99} is $r=0.05$~cm
and nothing can be said about distances closer to the source.

\begin{figure}[b]
\begin{center}
\parbox[c]{15cm}
{\includegraphics[scale=0.6]{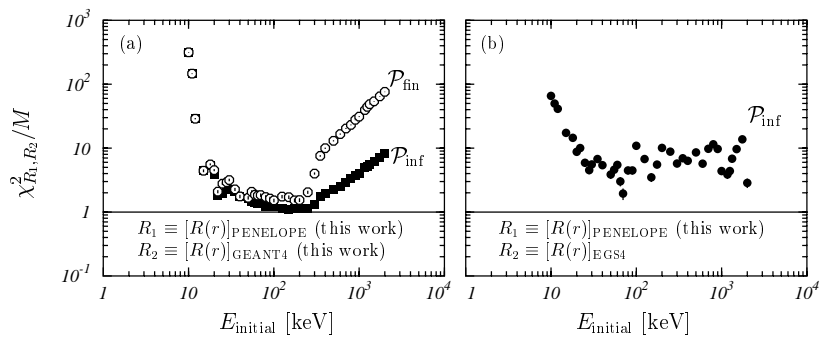}}
\end{center}
\vspace*{-.5cm}
\caption{\small Values of $\chi^2_{R_1,R_2}/M$, as given by
  Eq. (\ref{eq:chi2statistic}), calculated for the quantity $R(r)$, as
  a function of the energy of the initial photons. The comparison
  between the results of our simulations with PENELOPE and GEANT4 are
  shown for the ${\mathcal P}_{\rm inf}$ (black squares) and
  ${\mathcal P}_{\rm fin}$ (open circles) phantoms in panel (a). In
  panel (b), we show the results for the comparison of our PENELOPE
  results with those of \citet{Lux99} obtained with EGS4.
\label{fig:comp2}}
\end{figure}

The degree of agreement between the different calculations can be
evaluated by using the $\chi^2$ statistic defined in
Eq. (\ref{eq:chi2statistic}). In panel (a) of Fig. \ref{fig:comp2} we
show with black squares the results obtained for the statistic when
comparing the PENELOPE and GEANT4 results for the ${\mathcal P}_{\rm inf}$
phantom (plotted in left panels of Fig. \ref{fig:comp1}). The $\chi^2/M$
appears to be rather large at low energies, close to 1 at medium
energies and grows with energy above 200~keV. In these calculations of
the $\chi^2$ statistic, the full histograms have been considered. In
all cases the probability $Q(\chi^2 | M)$ is smaller than 0.001.

In the same panel we have plotted the $\chi^2/M$ values obtained for
the same comparison but using the ${\mathcal P}_{\rm fin}$
phantom. These values are practically the same as those previously
discussed for energies below 50~keV. The reason for this behavior is
that the finite phantom used is, in practice, infinite for
these low energies. At higher energies, the $\chi^2/M$ values are larger
for this finite phantom than for the infinite one. This is mainly due
to the fact that the number of bins included in the calculations is
smaller (roughly a factor 10) for ${\mathcal P}_{\rm fin}$: though the
differences between PENELOPE and GEANT4 are similar for both phantoms,
their relative importance is rather larger in the ${\mathcal P}_{\rm
fin}$ case.

In panel (b) of Fig. \ref{fig:comp2}, the results of the comparison
between our PENELOPE simulations for the ${\mathcal P}_{\rm inf}$ phantom
and the EGS4 calculations of \citet{Lux99} are
plotted. At low energy, the trend is similar to the one observed in
the previous panel. In this case, values near $\chi^2/M \sim 1$ are
not obtained. However, above 40~keV, the $\chi^2/M$ does not show the
quick growth with energy found in the comparison between PENELOPE and
GEANT4.

We have performed a comparison of our results with those of
\citet{Ye04} obtaining, as expected, a good agreement above 10~keV for
the phantom ${\mathcal P}_{\rm fin}$). For 10 keV a certain discrepancy is
observed which can be ascribed to the fact that these authors gave
results for five bins only.

Apart from the random variability, we have not found any particular
structure at 3~cm from the 2~MeV source, neither with PENELOPE nor
with GEANT4 using both the ${\mathcal P}_{\rm fin}$ and the ${\mathcal
P}_{\rm inf}$ phantoms.  This supports the guess of \citet{Lux99} and
their observation was, probably, a simulation artifact.

The differences observed between the different calculations can be due
to a variety of factors. For energies above a few hundred of keV, the
disagreement found close to the source between PENELOPE and GEANT4 can
be ascribed to the known problems of this last code with the multiple
scattering implementation \citep{Poo05b}, which affects the electron
transport and is crucial in the absence of charged particle
equilibrium, as it occurs precisely nearby the source. This is an
important point because, as mentioned above, we have done calculations
with GEANT4 using electron tracking parameters more exigent and we
have found the same results.

For low energies, below 80~keV, electron transport is negligible and,
at least in principle, the differences can be due only to the fact
that the photoelectric and Compton cross sections used by the
low-energy package of GEANT4 and PENELOPE are different. This
discrepancy should be reduced by using the package PENELOPE of GEANT4,
in which the same cross sections as in the PENELOPE code are used. As
shown by \citet{Poo05b}, the differences obtained for the
photoelectric mass attenuation coefficient using the PENELOPE package
and the low-energy package of GEANT4 are negligible. On the other
hand, the differences for the Compton mass attenuation coefficient are
smaller than 1.5\% for energies below 200~keV. Above this energy, no
differences are found. In order to see if these differences solve the
discrepancies shown by our calculations, we have performed simulations
using the PENELOPE package of GEANT4 for energies below 80~keV. First,
we have found that these new results differ from those corresponding
to the low-energy package by less than 1\% of the maximum of $R(r)$
and, second, the modification does not improve the agreement with the
results obtained with the PENELOPE code.

The differences observed between PENELOPE and EGS4 at low energies,
where electron transport is not relevant, can be ascribed to the
different cross sections used for photoelectric and Compton processes
in both codes. Above 80 keV, also the differences in the electron
tracking procedures used in these two codes affect. One can expect that
EGSnrc \citep{Kaw03}, a newer version of EGS, with new implementations of
the photoelectric and Compton process physics would reduce these 
differences to a large extent.

\section{Conclusions}

In this work, Monte Carlo codes PENELOPE and GEANT4 have been used to
calculate the dose rate in water for monoenergetic photon point
sources with energies of interest in brachytherapy, ranging between
10~keV and 2~MeV. 

To compare the results obtained with these two codes between them and
with those calculated by other authors, we have proposed a statistical
test based on the evaluation of the $\chi^2$ between two histograms. 

We have found some small discrepancies between the versions 2001 and
2003 of PENELOPE for source energies larger than 1~MeV, while the
versions 2005 and 2006 provide results in perfect agreement with version
2003 in the cases here analyzed.

The comparison between PENELOPE and GEANT4 results shows a reasonable
agreement for all energies analyzed except in the neighbor of the
source, up to a few millimeters.

In the case of the infinite phantom, the PENELOPE results are also in
reasonable agreement with previous EGS4 calculations of \citet{Lux99}
for distances larger than 1~cm.

The comparison of our results with the results of \citet{Ye04} shows a
good agreement above 10~keV for the finite phantom.

We have shown that the comparison between the histograms obtained as
output of this kind of simulations can produce contradictory results
if it is done on the basis of a restricted set of bins. To avoid this
problem, at least in part, we have proposed a $\chi^2$ test which
gives a quantitative estimate of the agreement or disagreement between
histograms. However, due to its integral character, other relevant
information can be lost (e.g., the possible bias, the bins which are
actually different, etc.) Thus, further investigation to find
additional statistics should be of interest in order to complete the
analysis.

\section{ACKNOWLEDGMENTS}
{This work has been supported in part by the Junta de
Andaluc\'{\i}a (FQM0220) and by the Ministerio de Educaci\'on y
Ciencia (FIS2005-03577). F.M.O. A.-D. acknowledges the University of
Granada and the Departamento de F\'{\i}sica Moderna for partially
funding his stay in Granada (Spain).}

\end{document}